\documentclass[sn-mathphys,Numbered]{sn-jnl}

\usepackage{graphicx}%
\usepackage{multirow}%
\usepackage{amsmath,amssymb,amsfonts}%
\usepackage{amsthm}%
\usepackage{mathrsfs}%
\usepackage[title]{appendix}%
\usepackage{xcolor}%
\usepackage{textcomp}%
\usepackage{manyfoot}%
\usepackage{booktabs}%
\usepackage{algorithm}%
\usepackage{algorithmicx}%
\usepackage{algpseudocode}%
\usepackage{listings}%
\usepackage{subfig}
\usepackage{float}

\theoremstyle{thmstyleone}%
\theoremstyle{thmstyletwo}%

\theoremstyle{thmstylethree}%

\begin{document}

\title[Article Title]{Utilizing Layout Effects for Analog Logic Locking}


\author[1]{\fnm{Muayad J.} \sur{Aljafar}}\email{muayad.al-jafar@taltech.ee}

\author[2]{\fnm{Florence} \sur{Azais}}\email{florence.azais@lirmm.fr}

\author[2]{\fnm{Marie-Lise} \sur{Flottes}}\email{marie-lise.flottes@lirmm.fr}

\author*[3,1]{\fnm{Samuel} \sur{Pagliarini}}\email{pagliarini@cmu.edu}

\affil[1]{\orgdiv{Dpt. of Computer Systems}, \orgname{Tallinn University of Technology}, \orgaddress{
\country{Estonia}}}

\affil[2]{\orgdiv{LIRMM}, \orgname{Université de Montpellier (UM) - CNRS}, \orgaddress{
\country{France}}}

\affil[3]{\orgdiv{Electrical and Computer Eng.}, \orgname{Carnegie Mellon University}, \orgaddress{
\country{USA}}}


\abstract{While numerous obfuscation techniques are available for securing digital assets in the digital domain, there has been a notable lack of focus on protecting Intellectual Property (IP) in the analog domain. This is primarily due to the relatively smaller footprint of analog components within an Integrated Circuit (IC), with the majority of the surface dedicated to digital elements. However, despite their smaller nature, analog components are highly valuable IP and warrant effective protection. In this paper, we present a groundbreaking method for safeguarding analog IP by harnessing layout-based effects that are typically considered undesirable in IC design. Specifically, we exploit the impact of Length of Oxide Diffusion and Well Proximity Effect on transistors to fine-tune critical parameters such as transconductance ($g_{m}$) and threshold voltage ($V_{th}$). These parameters remain concealed behind key inputs, akin to the logic locking approach employed in digital ICs. Our research explores the application of layout-based effects in two commercial CMOS technologies, namely a 28nm and a 65nm node. To demonstrate the efficacy of our proposed technique, we implement it for locking an Operational Transconductance Amplifier. Extensive simulations are performed, evaluating the obfuscation strength by applying a large number of key sets (over 50,000 and 300,000). The results exhibit a significant degradation in performance metrics, such as open-loop gain (up to 130dB), phase margin (up to 50 degrees), 3dB bandwidth (approximately 2.5MHz), and power consumption (around 1mW) when incorrect keys are employed. Our findings highlight the advantages of our approach as well as the associated overhead.}

\keywords{Analog Obfuscation, Layout-based effects, Logic Locking, Hardware Security}



\maketitle

\section{Introduction}\label{sec1}

The outsourcing of fabrication in the semiconductor supply chain has exposed it to numerous security threats, including Integrated Circuit (IC) piracy, counterfeiting, overproduction, and hardware Trojans \cite{b1,b2,b3,b4}. These threats have resulted in significant annual losses, estimated at \$4 billion a decade ago \cite{b5}. To mitigate these security risks, design-for-trust (DfTr) techniques have been developed, primarily focused on digital ICs \cite{b6,b7}. One prominent example of a DfTr technique is logic locking \cite{b8}.

However, the research efforts to secure analog ICs or analog Intellectual Property (IP) have been relatively limited. Analog ICs are susceptible to security threats due to their small footprint and widespread use across various application domains. In fact, pirating analog ICs, which typically consist of a few hundred transistors, is often easier compared to digital ICs with millions of transistors. It should also be noted that in digital design the transistor are often sized with the minimum allowed length and width parameters, which is not necessarily true for analog designs. Previous studies on analog logic locking have explored techniques such as key provisioning \cite{b9} and tuning circuit functionalities \cite{b10}, involving the concealment of voltage or current biases, transistor sizing, or voltage thresholds of devices \cite{b11,b12,b13,b14,b15,b16,b17,b28}. Additionally, some techniques have been applied to lock the digital portion of Analog Mixed-Signal (AMS) circuits using digital logic locking methods \cite{b18,b19}. Vulnerability assessments of obfuscated analog circuits have been conducted \cite{b20}, and attacks utilizing Satisfiability Modulo Theories (SMT), bias locking, and genetic algorithm have been proposed \cite{b21,b22, b27}. However, the approach presented in this paper, which is an extension of our previous research work \cite{b29}, introduces a completely novel method for analog obfuscation by leveraging layout-based effects to establish a key-based lock. Our method is the first to utilize this unconventional approach.

\begin{figure}[b]
\centerline{\includegraphics[width=0.98\linewidth]{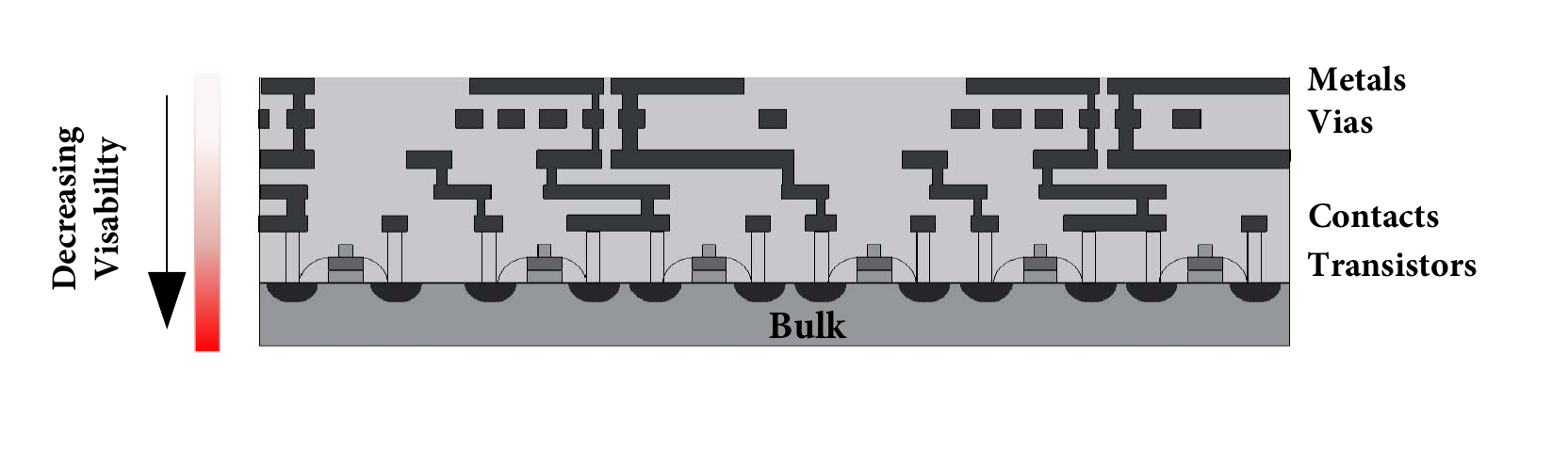}}
\caption{Cross section of the metal stack in an IC. Moving towards the base layer, the visibility decreases and the effort to reverse engineer increases.}
\label{IC_metal_stack}
\end{figure}

We propose a novel technique for securing analog ICs against counterfeiting and Reverse Engineering (RE) attacks, which aim to clone ICs or extract proprietary information such as netlists and layouts. Counterfeiting involves selling cloned or illegitimately overproduced ICs in the aftermarket, while RE attacks aim to derive confidential information from ICs. In RE attacks, the adversary undergoes a process of depackaging the IC, delayering it, capturing images of the layers, and reconstructing the netlist using specialized image processing tools. While this process has its challenges, such as handling a large number of images, it still succesfully reveals the metal lines, vias, and contacts. However, as the delayering process approaches the transistor layers, the features become increasingly difficult to obtain. Obtaining low-level properties like doping gradients at the device level solely through delayering and imaging is non-trivial. Figure \ref{IC_metal_stack} illustrates the complexity involved in RE of a complex metal stack.
In this research, we introduce obfuscation by manipulating two low-level properties in the diffusion layer, specifically the \textit{Well Proximity Effect} (WPE) and the \textit{Length of Diffusion} (LOD). These properties, also known as local layout effects or Layout-Dependent Effects (LDEs), are challenging to identify or characterize compared to transistor size. To date, no RE attack has demonstrated the capability to extract this level of detail, and the process of obtaining such information is deemed costly and time-consuming \cite{b23}. However, these effects directly impact transistor behavior, including parameters such as threshold voltage ($V_{th}$) and transconductance ($g_{m}$), which in turn affect the performance of analog circuits. For instance, in an Operational Transconductance Amplifier (OTA), these effects would influence power consumption, gain, phase, and transconductance parameters.

This work presents several significant contributions, which are outlined below:
\begin{itemize}
\item Introduction of a novel approach: The paper demonstrates, for the first time, how to leverage undesirable layout-based effects to effectively lock analog circuits. This innovative technique adds a new dimension to analog circuit protection.
\item Scalability and adaptability of the proposed technique across different process technologies: The validation results obtained from 28nm and 65nm technology nodes confirm that the locking mechanism can be implemented effectively in various manufacturing processes, enhancing the security of analog circuits in different technology generations.
\item Demonstrating the deterministic nature of LDEs, even in the presence of parasitics and process variation.
\end{itemize}

The remaining sections of the paper are organized as follows: Section \ref{sec2} introduces and explains the proposed technique in detail. Section \ref{sec3} presents a comprehensive case study, demonstrating the application of the locking technique and providing the corresponding results. Section \ref{sec4} discusses potential attack models and conducts a security analysis. Finally, Section \ref{sec5} concludes the paper, summarizing the findings and emphasizing the contributions of this research.

\section{Background and Proposed Locking Technique}\label{sec2}

\subsection{Layout-Dependent Effects}\label{subsec21}

\begin{figure}[t]
\centerline{\includegraphics [width=.55\linewidth]{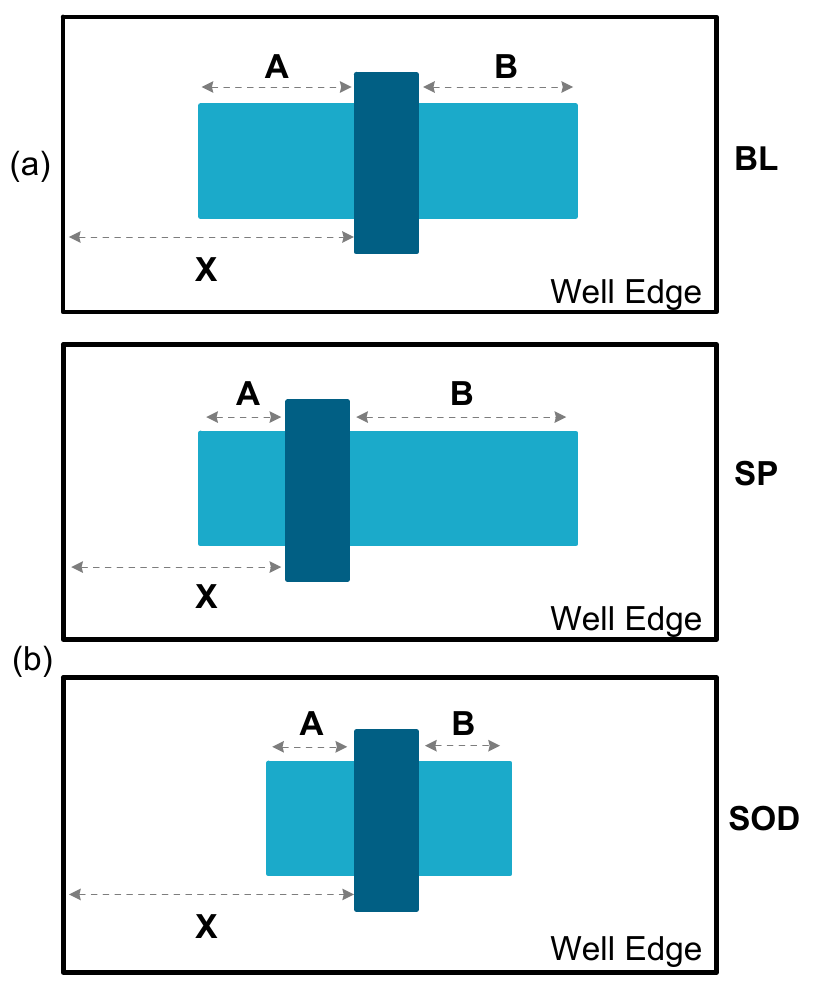}}
\caption{Layout-dependent effects. (a) Simplified transistor layout, baseline. (b) Different transistor arrangements, used for obfuscating analog circuits. OD is the oxide diffusion, and PO is the poly (gate). A and B are the distances between the poly and the OD edges, and X is the distance between the poly and the well edges. X relates to WPE, and A and B relate to LOD.}
\label{transistor_layouts}
\end{figure}

Layout-dependent effects emerge as a consequence of the reduction in process geometries during lithography. Among these effects in sub-100nm CMOS technologies, it is known that the electrical behavior of a device (i.e., a transistor) depends on its well proximity and on its length of diffusion. However, it is important to note that WPE and LOD are not the only LDE effects that exist. More advanced nodes have many other effects such as poly and poly-cut related issues. 

WPE is closely related to the proximity of a device to the well edge. Transistors located near the well edge exhibit different performance characteristics, such as voltage threshold and drain current, compared to those positioned farther from the well edge (represented as X in Figure \ref{transistor_layouts}). This discrepancy arises from the scattering of implant ions off the resist side-well, even when the transistors have identical dimensions.
LOD, on the other hand, arises from the mechanical stress induced by different lengths of oxide (illustrated as A and B in Figure \ref{transistor_layouts}). These variations in OD length affect carrier mobility, thereby impacting the current flow within the devices.
 
\begin{figure}[t]
\centering
\includegraphics[width=.65\linewidth]{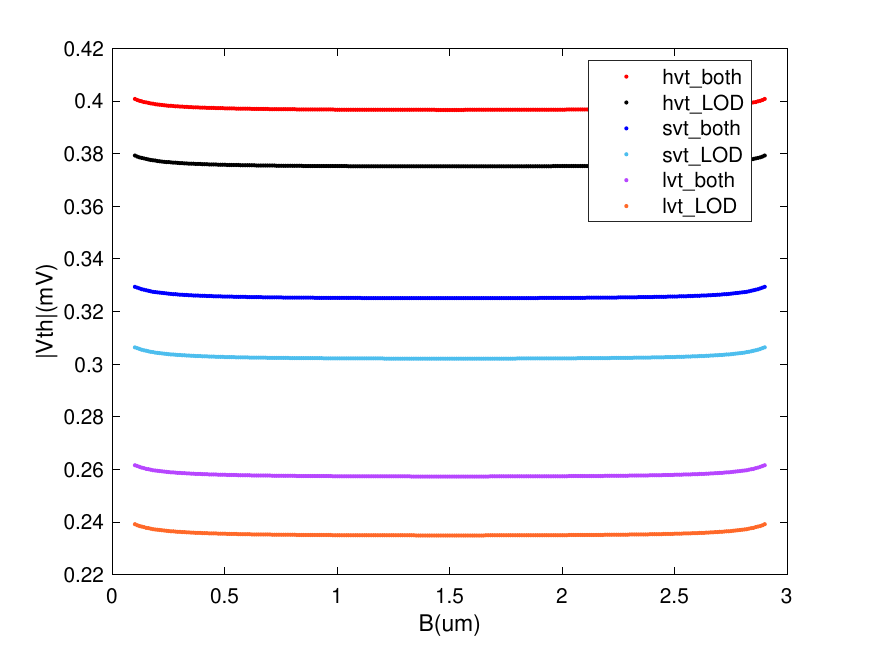}
\includegraphics[width=.65\linewidth]{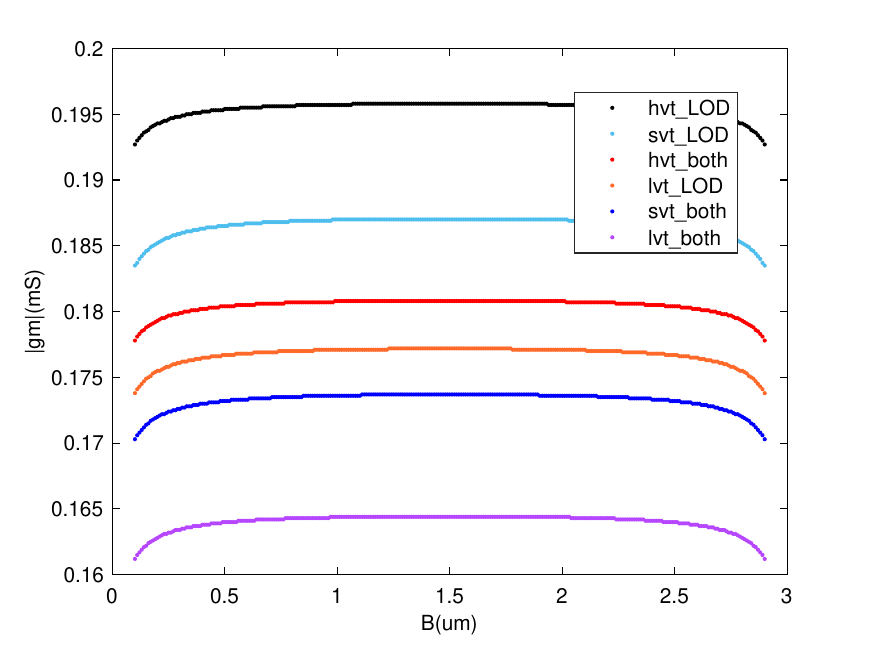}
\caption{Effects of LOD and both LOD and WPE on the absolute values of voltage threshold and transconductance of PMOS transistors with a minimum length and representative width. B is shown in Figure \ref{transistor_layouts}.}
\label{vth_gm_LDEs}
\end{figure}

Figure \ref{vth_gm_LDEs} illustrates the impact of LOD and the combined effects of LOD and WPE on the absolute values of voltage threshold and transconductance for a 65nm PMOS transistor with standard (SVT-), high (HVT-), and low (LVT-) voltage thresholds at a $V_{gs}$ of 1V. 
In Figure \ref{vth_gm_LDEs}, when the value of B (as shown in Figure \ref{transistor_layouts}) is very small or very large, indicating that the poly is in close proximity to the sides of the OD, the transistor exhibits distinct $V_{th}$ and $g_m$ values compared to other B values. This observation forms the basis of leveraging layout-dependent effects for the obfuscation of analog ICs in this study. 
These layout-dependent effects have a similar impact on the performance of an NMOS transistor and contribute to device mismatch in analog circuits.

In this study, our objective is to leverage these layout effects to implement a locking mechanism for analog circuits. We consider three different \textbf{arrangements or configurations} for a transistor: baseline (BL), side-poly (SP), and short-OD (SOD), as illustrated in Figure \ref{transistor_layouts}. The baseline configuration represents the nominal case of layout-dependent effects, while SP and SOD configurations are utilized to further exploit WPE and LOD effects.
By employing these different arrangements, we can achieve variations of approximately 10\% in voltage threshold and transconductance compared to the baseline case. The magnitude of voltage threshold variations is larger for NMOS transistors compared to PMOS transistors, whereas the transconductance variations are smaller for NMOS transistors compared to PMOS transistors (Table \ref{variations_vth_gm}). Statistical variations due to both process variations and mismatch were also simulated for all configurations. Table \ref{process_mismatch} presents the standard deviations (SD) of $V_{th}$ and $gm$ with respect to their mean values.
The results reported in Table \ref{variations_vth_gm} and Table \ref{process_mismatch} demonstrate the deterministic nature of layout-based effects. Regardless of where the fabricated IC falls within the process variation spectrum, these effects consistently manifest themselves.

\begin{figure}[]
\centerline{\includegraphics[width=.65\linewidth]{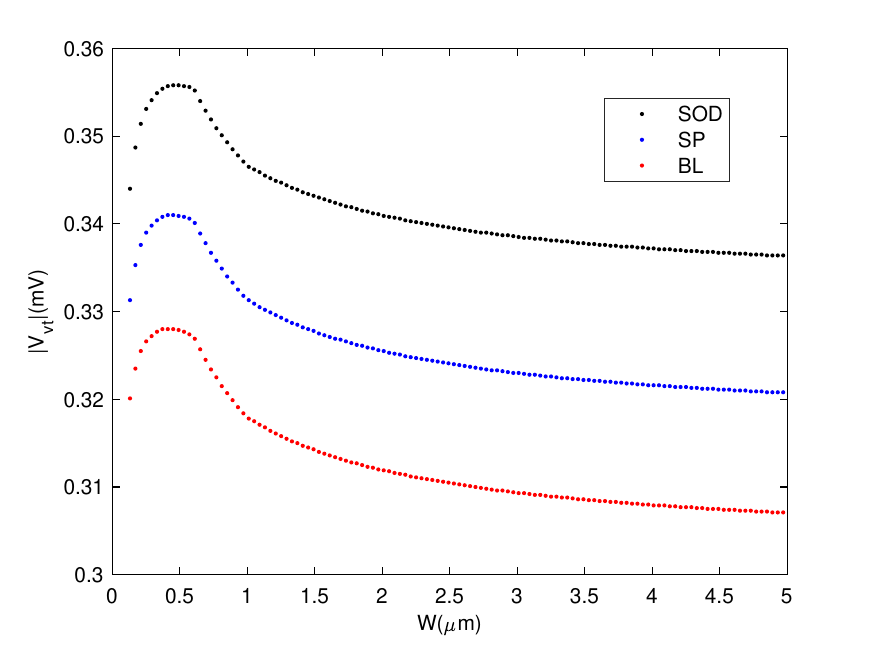}}
\caption{Effects of PMOS width, WPE and LOD on the absolute values of voltage thresholds for all arrangements.}
\label{length-vth_LDEs}
\end{figure}
Figure \ref{length-vth_LDEs} illustrates the impact of transistor width ($W$) in conjunction with the layout-dependent effects on the voltage thresholds for all transistor arrangements. In this plot, PMOS transistors with minimum length are considered. It is noteworthy that the margin between the lines representing the BL-SP and BL-SOD configurations remains nearly constant, indicating that transistors of any size can be potentially used for obfuscation. In the specific example shown, increasing the $W$ in the SOD configuration leads to a change in voltage threshold variations from $6.8\%$ to $8.7\%$ compared to the BL configuration. This demonstrates that altering the transistor width can further enhance the effectiveness of the layout-based effects for obfuscation purposes.

\begin{table}[]
	\caption{Variations (\%) in Voltage Threshold and Transconductance with respect to BL (the baseline)}\label{variations_vth_gm}
		\begin{tabular}{c c c c c c c}
			\hline
			\hline
            \multirow{2}{*}{} &         					 & \multirow{2}{*}{$A_i$}   & \multirow{2}{*}{\sc Device}  &  \multicolumn{3}{c}{\sc Variations}\\\cmidrule{5-7}

	                         &          			        &        				   &                              & {\sc hvt(\%)} &{\sc svt(\%)} & {\sc lvt(\%)}\\ \hline
	        
            \parbox[t]{2mm}{\multirow{12}{*}{\rotatebox[origin=c]{90}{\sc parameter}}}       
                              & \parbox[t]{2mm}{\multirow{5}{*}{\rotatebox[origin=c]{90}{\sc $V_{th}$}}} 
                                                            & \multirow{2}{*}{\sc SP}   & {\sc pmos}                   & {\sc 2.85}    & {\sc 3.7}     & {\sc 4.59} \\\cmidrule{4-7}

			                 &          			       &         				   & {\sc nmos}                   & {\sc 4.05}    & {\sc 4.38}    & {\sc 5} \\\cmidrule{4-7}
			                 &        			            & \multirow{2}{*}{\sc SOD} 	& {\sc pmos}                   & {\sc 6.08}   & {\sc 7.9}     & {\sc 9.79}  \\\cmidrule{4-7}
			                 &          				    &        				    & {\sc nmos}                   & {\sc 8.53}   & {\sc 9.28}    & {\sc 10.61}    \\\cmidrule{2-7}
                             & \parbox[t]{2mm}{\multirow{5}{*}{\rotatebox[origin=c]{90}{\sc $gm$}}}  	 
                                                            & \multirow{2}{*}{\sc SP}  & {\sc pmos}                   & {\sc 4.76}    & {\sc 4.72 }   & {\sc 4.68}    \\\cmidrule{4-7}
			                 &          				    &         					& {\sc nmos}                   & {\sc 1.72}    & {\sc 2.54}    & {\sc 2.42}    \\\cmidrule{4-7}
		                     &            		            & \multirow{2}{*}{\sc SOD} 	& {\sc pmos}                   & {\sc 10.4}    & {\sc 10.19 }  & {\sc 10.16}   \\\cmidrule{4-7}
	     	                 &           		            &        					& {\sc nmos}                   & {\sc 3.7}     & {\sc 5.41}    & {\sc 5.09}     \\
			\hline
			\hline						
		\end{tabular}	
	\small{Values were obtained from corner analysis for typical corner for devices with a minimum length and representative width. $A_i$ is an arrangement as defined in Figure \ref{transistor_layouts}}
\end{table}

\begin{table}[]
	\caption{Process and mismatch of the arrangements}\label{process_mismatch}
	\begin{center}
		\begin{tabular}{c c c c c c c}
			\hline
			\hline
			\multirow{2}{*}{}   &          & \multirow{2}{*}{$A_i$}   & \multirow{2}{*}{\sc Device}  &  \multicolumn{3}{c}{\sc Variations in SD*} \\\cmidrule{5-7}
		                      &      	 &        				    &                              & {\sc hvt(\%)} & {\sc svt(\%)} & {\sc lvt(\%)}  \\
			\hline
			\parbox[t]{2mm}{\multirow{18}{*}{\rotatebox[origin=c]{90}{\centerline{\sc parameter}}}}       
			                  & \parbox[t]{2mm}{\multirow{8}{*}{\rotatebox[origin=c]{90}{\sc $V_{th}$}}}   
			                             & \multirow{2}{*}{\sc BL} 	 & {\sc pmos}                   & {\sc 9.78}     & {\sc 10.64}   & {\sc 12.95}   \\ \cmidrule{4-7}
			                  &          &         					& {\sc nmos}                   & {\sc 15.34}    & {\sc 12.29}   & {\sc 9.73}    \\ \cmidrule{4-7}
			                  &        	 & \multirow{2}{*}{\sc SP} 	& {\sc pmos}                   & {\sc 9.38}     & {\sc 10.12}   & {\sc 12.17}   \\ \cmidrule{4-7}
		                      &          &        				    & {\sc nmos}                   & {\sc 14.07}    & {\sc 11.28}   & {\sc 9.16}    \\ \cmidrule{4-7}
		                      &     	 & \multirow{2}{*}{\sc SOD}	& {\sc pmos}                   & {\sc 8.97}     & {\sc 9.58}    & {\sc 11.37}   \\ \cmidrule{4-7}
		                      &      	 &        				    & {\sc nmos}                   & {\sc 12.88}    & {\sc 10.34}   & {\sc 8.61}    \\ \cmidrule{2-7}
		                      & \parbox[t]{2mm}{\multirow{8}{*}{\rotatebox[origin=c]{90}{\sc $gm$}}}	 
			                             & \multirow{2}{*}{\sc BL}	 & {\sc pmos}                   & {\sc 3.55}     & {\sc 3.98 }   & {\sc 2.90}    \\ \cmidrule{4-7}
		                      &      	 &         					& {\sc nmos}                   & {\sc 3.78}     & {\sc 2.85}    & {\sc 5.93}    \\\cmidrule{4-7}
			                  &          & \multirow{2}{*}{\sc SP}	& {\sc pmos}                   & {\sc 3.35}     & {\sc 3.94 }   & {\sc 2.83}   \\ \cmidrule{4-7}
			                  &    		 &        					& {\sc nmos}                   & {\sc 3.63}     & {\sc 2.84}    & {\sc 5.98}     \\ \cmidrule{4-7}
		                      &    		 & \multirow{2}{*}{\sc SOD}	& {\sc pmos}                   & {\sc 3.17}     & {\sc 3.91 }   & {\sc 2.75}   \\ \cmidrule{4-7}
		                      &          &        					& {\sc nmos}                   & {\sc 3.45}     & {\sc 2.85}    & {\sc 6.05}     \\
			\hline
			\hline
			
		\end{tabular}	
		\vspace{2pt}
		\item \small {\centering *SD means standard deviation.}		 
	\end{center}
\end{table}

\subsection{LDEs in Sub-100nm Technologies}\label{subsec22}

\begin{figure}[t]
\centerline{\includegraphics[width=.75\linewidth]{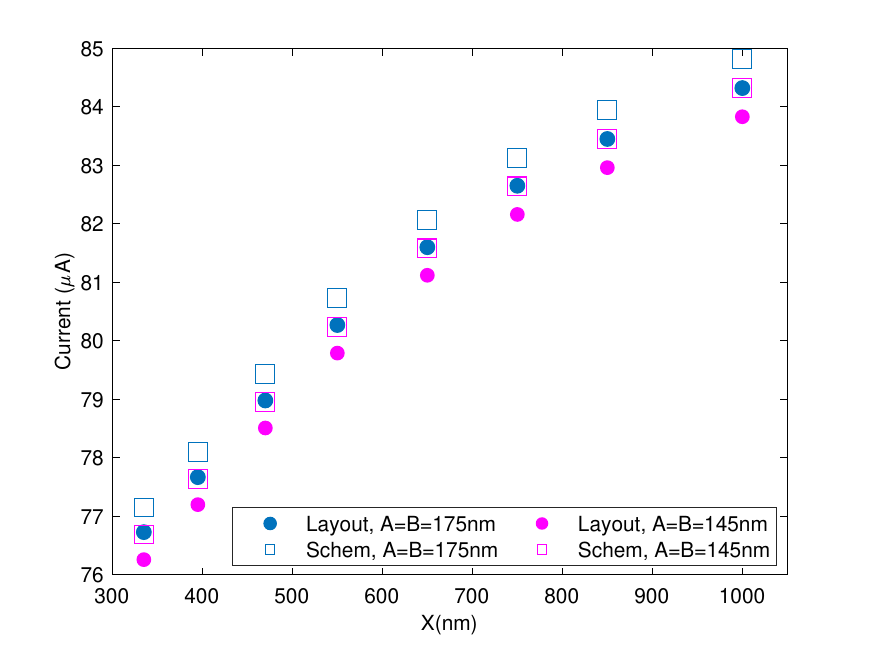}}
\caption{Impact of layout-dependent effects on an LVT PMOS drain current in 65nm technology.}
\label{LDEs_schematicLayout_65nm}
\end{figure}

\begin{figure}[H]
\centerline{\includegraphics[width=.75\linewidth]{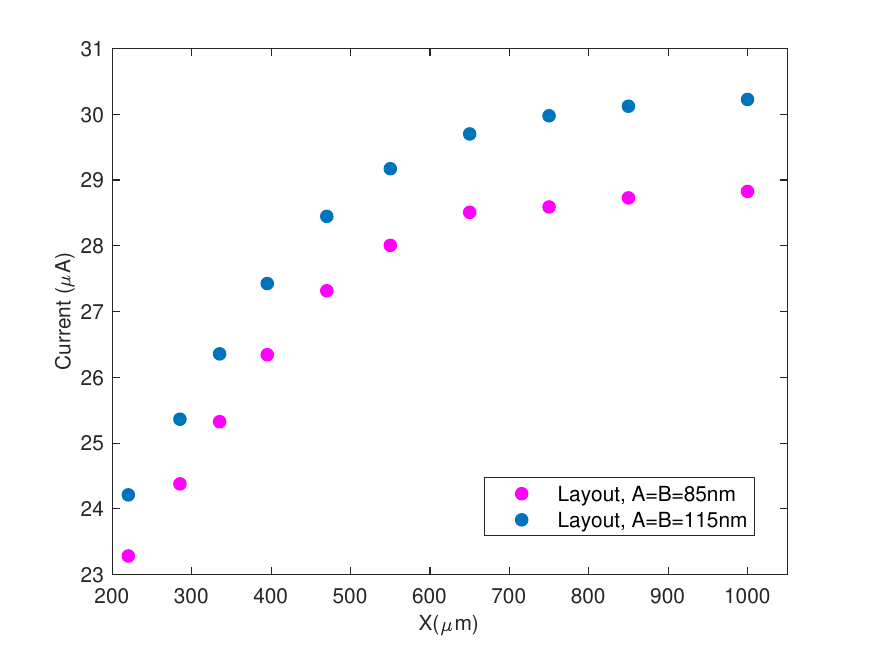}}
\caption{Impact of layout-dependent effects on an LVT PMOS drain current in 28nm technology.}
\label{LDEs_28nm}
\end{figure}

We now demonstrate the influence of well proximity effects and length of diffusion on key parameters of transistors in both 28nm and 65nm technologies, considering parasitic effects. Additionally, we highlight the impact of these LDEs on digital circuits, specifically on the behavior of CMOS inverters.
Figure \ref{LDEs_schematicLayout_65nm} presents the drain current variations of an LVT PMOS transistor in 65nm technology due to LDEs. Schematic and layout simulations are conducted for different values of A, B, X, and an applied $v_{gs}$ of 1V. The results indicate a consistent trendline, showing an increase in drain current with higher values of X (and A and B). Similarly, Figure \ref{LDEs_28nm} illustrates the impact of LDEs on the drain current of an LVT PMOS transistor in 28nm technology. Layout simulations are performed for different values of X and an applied $v_{gs}$ of 0.9V. The obtained results align with those observed in the 65nm technology simulations.

\begin{figure}[h]
\centering
\subfloat[Layouts]{
\centering
\includegraphics[width=.35\linewidth]{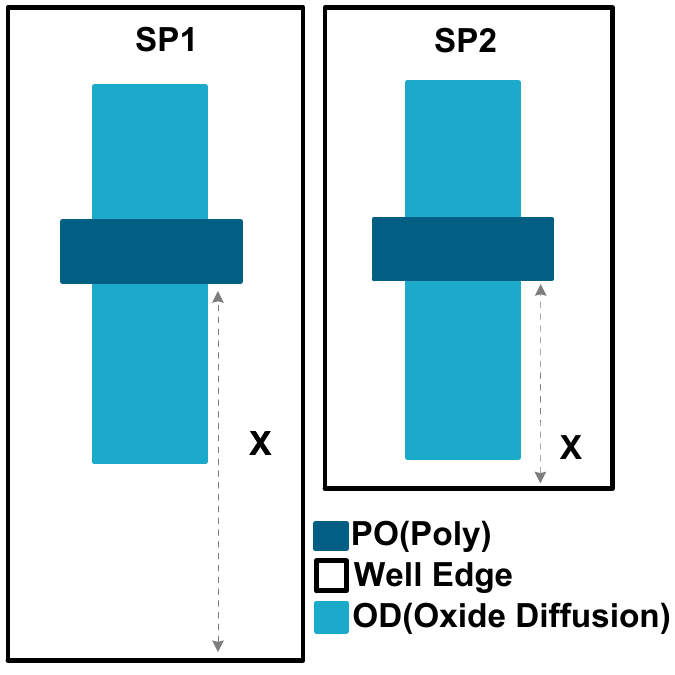}
}
\hfill
\subfloat[Transient Response]{
\centering
\includegraphics[width=.6\linewidth]{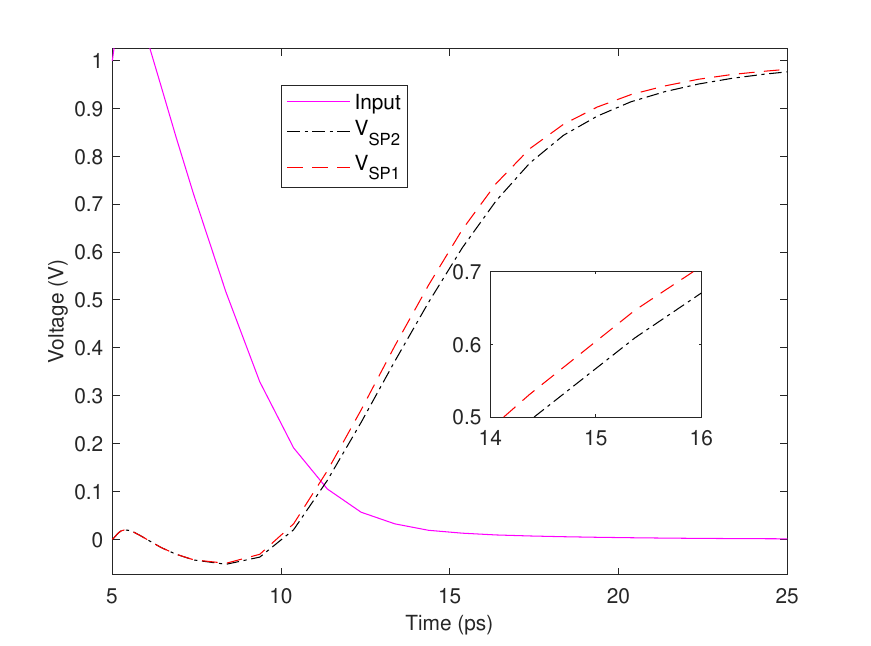}
}
\caption{Transistor layouts with different variants of the SP arrangement. The impact of WPEs on transient response of inverters with these layouts for PMOS transistors.}
\label{transientResponse}
\end{figure}

\begin{figure}[h]
\centerline{\includegraphics[width=.75\linewidth]{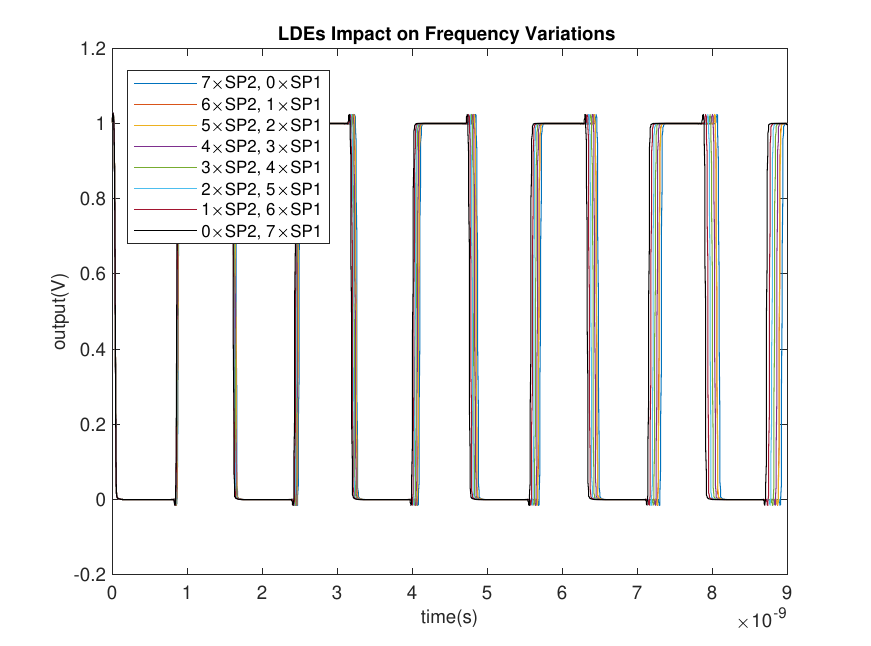}}
\caption{Impact of LDEs on a 7-stage ring oscillator’s frequency.}
\label{freq_variations}
\end{figure}

Furthermore, LDEs can impact the specifications of digital circuits, as demonstrated by their effect on the transient response of CMOS inverters.
In Figure \ref{transientResponse}, the transient response of two inverters with different variants of SP for PMOS transistors is shown. The inverter with the SP1 arrangement exhibits a slightly faster transient response ($V_{SP1}$) compared to the inverter with the SP2 arrangement ($V_{SP2}$). This difference in transient response indicates that the SP arrangements introduce variations in the propagation delay and rise/fall times of the inverters.
When these manipulated inverters are used to replace the original inverters in a 7-stage ring oscillator, the frequency of the oscillator deviates accordingly, as shown in Figure \ref{freq_variations}. Specifically, when inverters with the SP2 arrangement for PMOS transistors are utilized, the oscillator's frequency decreases. This decrease in frequency can be attributed to the slower transient response and longer propagation delay introduced by the SP2 arrangement.
On the other hand, replacing each of these inverters one by one with the variant inverters featuring the SP1 arrangement for PMOS transistors gradually increases the oscillator's frequency. This increase in frequency is due to the faster transient response and shorter propagation delay associated with the SP1 arrangement.
The observed frequency variations, in the range of a few MHz, highlight the significant impact of LDEs on the specifications of the 7-stage ring oscillator. These variations demonstrate that the layout configurations and arrangements of transistors can have a substantial influence on the performance of digital circuits.

\subsection {Applied Technique for Obfuscating Analog Circuits}\label{subsec23}
\begin{figure}[t]
\centerline{\includegraphics[width=.7\linewidth]{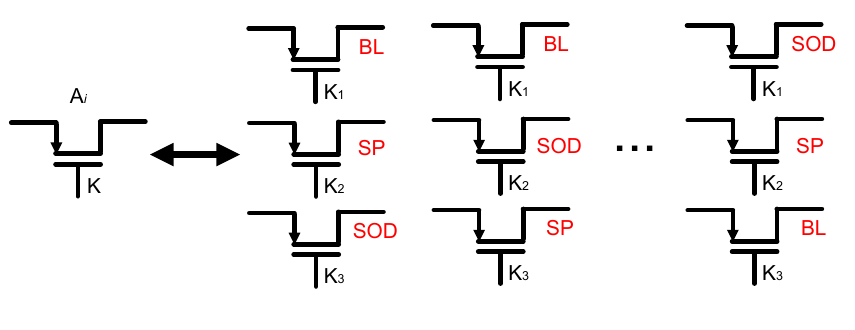}}
\caption{Principle of locking analog circuit. The order of these arrangements in the layout design is arbitrary. The figure shows only 3 out of 6 possible orders.}
\label{LL_principle}
\end{figure}

We propose a method for designing analog circuits using different arrangements of transistors, with the correct arrangement determined by a set of key bits. Each NMOS or PMOS transistor can have three possible arrangements, and the order of these arrangements in the layout can be arbitrary (Figure \ref{LL_principle}). Therefore, the correct key values correspond to a specific order of the arrangements. The key length for the entire circuit is determined by the number of devices, with each device requiring three key bits. This results in a total of $2^{3N}$ possible keys, assuming binary key signals.

However, it has been observed that some of the `wrong' keys can still result in desirable performance, while others may lead to nearly correct or completely incorrect behavior. To efficiently obfuscate an analog IP, we propose a simple three-step procedure:
\begin{enumerate}
\item Design the circuit using a combination of BL, SP, and SOD transistors.
\item Evaluate the impact of the two alternative arrangements for each transistor that were not originally employed.
\item Retain only the arrangements that result in incorrect performance, thereby promoting obfuscation.
\end{enumerate}
This three-step process can be enhanced by prioritizing certain configurations of transistors. Specifically, it is advantageous to convert transistors with multiple fingers into single-finger transistors whenever possible. This amplifies the performance shifts caused by layout-based effects. Additionally, exhaustive examination of all transistors is not necessary. Transistors can be randomly selected, and alternative arrangements can be chosen for evaluation. Circuit symmetry analysis and the designer's experience can be leveraged to identify a starting point for transistor selection.
Finally, the third step can be modified to discard arrangements that result in performance too close to the desired performance. If such ``undesirable’’ arrangements are identified, they can be eliminated. In Section \ref{sec3}, we provide a case study involving an OTA and implement the three-step procedure to lock the circuit.

\section{Case Study: Operational Transconductance Amplifier}\label{sec3}

\begin{figure}[t]
\centerline{\includegraphics[width=.8\linewidth]{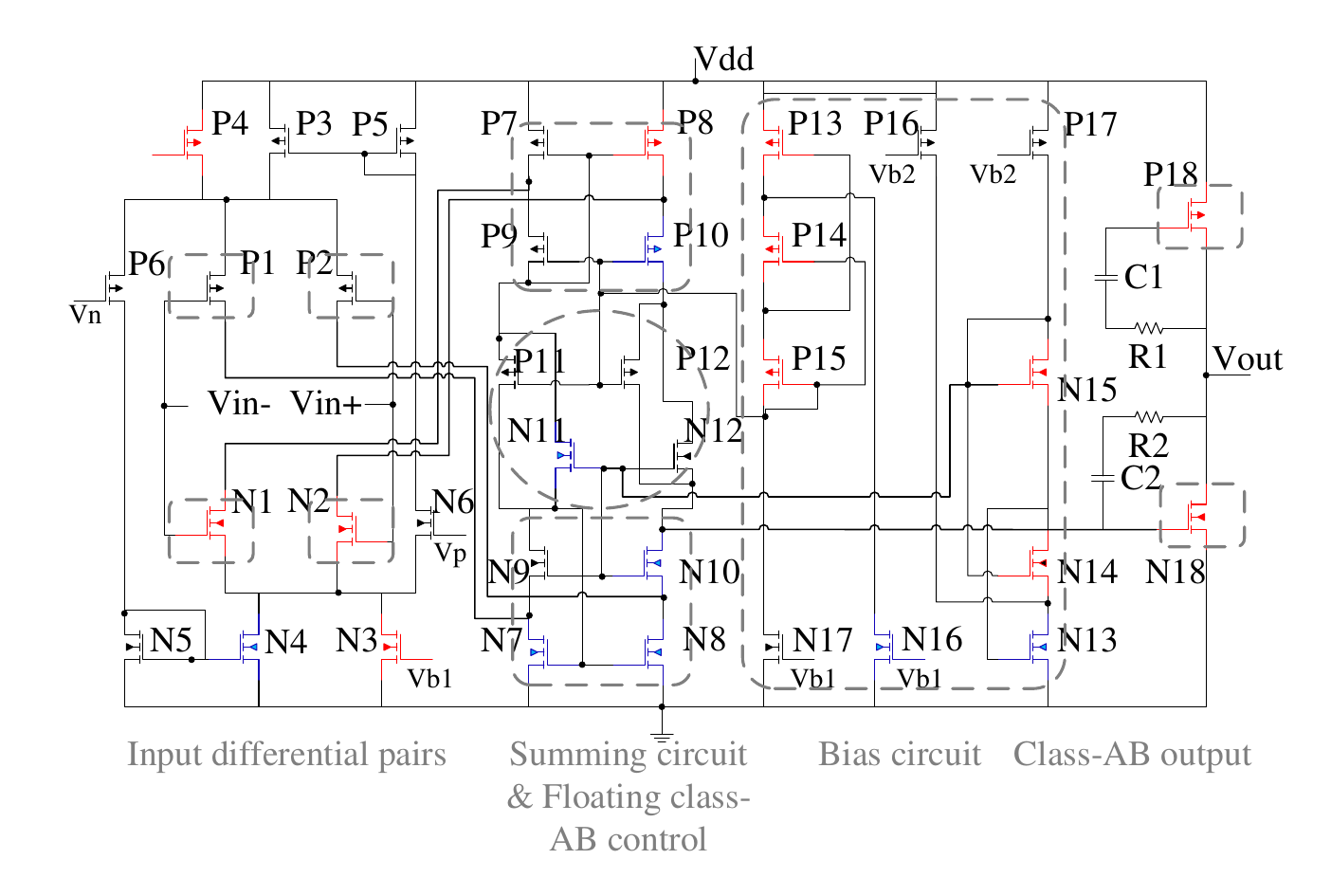}}
\caption{Schematic of OTA circuit. Multiple subcircuits such as input differential pairs (P1, P2, N1, N2), summing circuit (P7-P10, N7-N10), bias circuit (P13-P17, N13-N17), and class-AB output (P18, N18) are used for applying the layout-based effects. Red, blue, and black transistors represent arrangements SOD, SP, and BL, respectively.}
\label{OTA}
\end{figure}

We utilize the proposed technique to lock an operational transconductance amplifier as depicted in Figure \ref{OTA}. The specifications of the OTA for the chosen transistor arrangements are provided in Table ~\ref{specs_OTA}. It is important to note that, for this specific case study, we exclusively employ transistors with a standard voltage threshold. However, this does not imply a limitation of our technique, as it can be applied to transistors with different voltage thresholds as well.
Next, we investigate the impact of unused transistor arrangements on the performance of the OTA. The circuit comprises a total of 36 transistors, resulting in a search space of $2^{36 \times 3}$ possible arrangements. In practice, it is not feasible to examine all arrangements for every transistor. 
However, we can focus on those arrangements that are likely to affect the input differential pairs, summing circuit, floating class-AB control, bias block, and class-AB output, as indicated in Figure \ref{OTA}. This approach aligns with our earlier observation of leveraging the designer's expertise and considering circuit symmetry when selecting transistors for examination.

\begin{table}[h]
	\caption{\centering{OTA Specs for utilized arrangements in Figure \ref{OTA}}}\label{specs_OTA}
	\begin{center}
		\begin{tabular}{c c c c c}
			\hline
			\hline
			\multicolumn{5}{c}{\sc Specs}    \\
			\hline
			{\sc $gm$} &  {\sc Power*}  & {\sc Gain*} & {\sc Phase} & {3dB \sc Bandwidth}  \\
			\hline
			1.32mS  & 1.1mW   & 73.6dB  & 90deg  & 641KHz \\	
			\hline
			\hline			
		\end{tabular}		
		\item \centering *Power is the DC power, and gain is the open-loop gain.
		
	\end{center}	
\end{table}

\begin{figure}[t]
\centerline{\includegraphics[width=.6\linewidth]{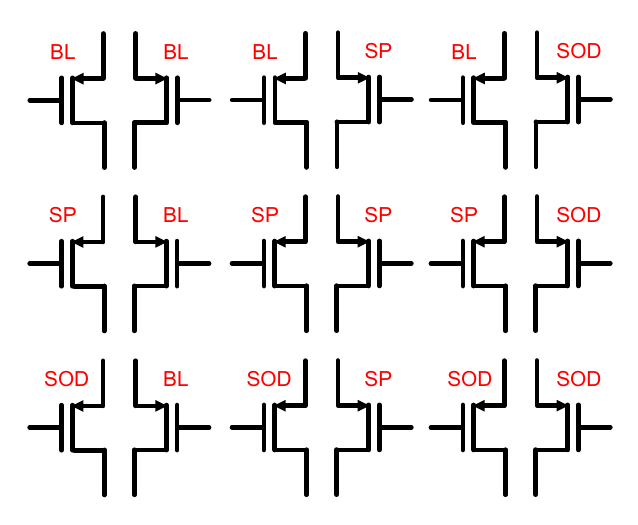}}
\caption{Pairs of arrangements. One pair of arrangements in the based design can be hidden among a subset of 8 pairs of arrangements.}\label{pairs_of_transistors}
\end{figure}

\emph{Simulation results}:  In our simulations, we utilize the Virtuoso Spectre circuit simulator in conjunction with a commercial 65nm technology. To initiate the obfuscation process, we select a set of 13 transistors from various parts within the circuit. These transistors are chosen arbitrarily and include P1, P2, P7, P8, P9, P10, N7, N8, N9, N10, N17, N18, and P18.
After selecting the 13 transistors for obfuscation, the initial keyspace consists of $2^{13\times3}$ possible keys. However, not all of these keys are suitable for effectively obfuscating circuit performance, so we apply our three-step procedure to improve the obfuscation at a cost of shrinking the keyspace. 
One important observation is that certain transistors can have varying degrees of impact on circuit performance and performance deviation caused by LDEs. To achieve a more balanced performance deviation, we have devised a strategy. Rather than obfuscating individual transistors, we choose to obfuscate a pair of transistors by tying together their select bits. This approach results in a more balanced LDE-induced performance deviation in a pair of transistors and enhances the overall effectiveness of the obfuscation technique.
\begin{figure}[t]
\centering
\subfloat[]{
\centering
\includegraphics[width=.48\linewidth]{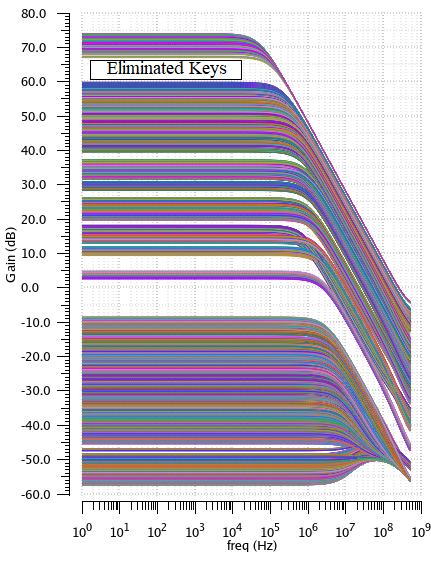}
}
\hfill
\subfloat[]{
\centering
\includegraphics[width=.48\linewidth]{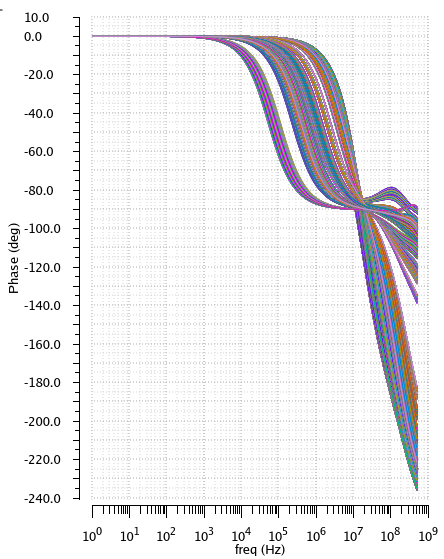}
}
\caption{Layout-based effects on (a) the OTA gain and (b) phase simulated for 50K keys. The gap marked in the graph is the result of purposefully removing nearly correct keys.}
\label{gain_phase}
\end{figure}
In the obfuscation process, each pair of transistors in the base design can be hidden among other pairs of arrangements, providing multiple possibilities for obfuscation (Figure \ref{pairs_of_transistors}). In this particular case, the 13 selected transistors for obfuscation form 6 pairs of transistors and one individual transistor. To obfuscate these 6 pairs, we introduce a different number of pairs of arrangements. Specifically, we add a total of $28$ random pairs of arrangements to hide the pairs of transistors, along with an additional single arrangement to obfuscate the individual transistor. Consequently, the key length for this experiment is $36$ bits (i.e., $28+6+1+1$), achieved by adding $57$ arrangements (i.e., $28\times2+1$) to the original design. 

\begin{figure}[t]
\centerline{\includegraphics[width=.75\linewidth]{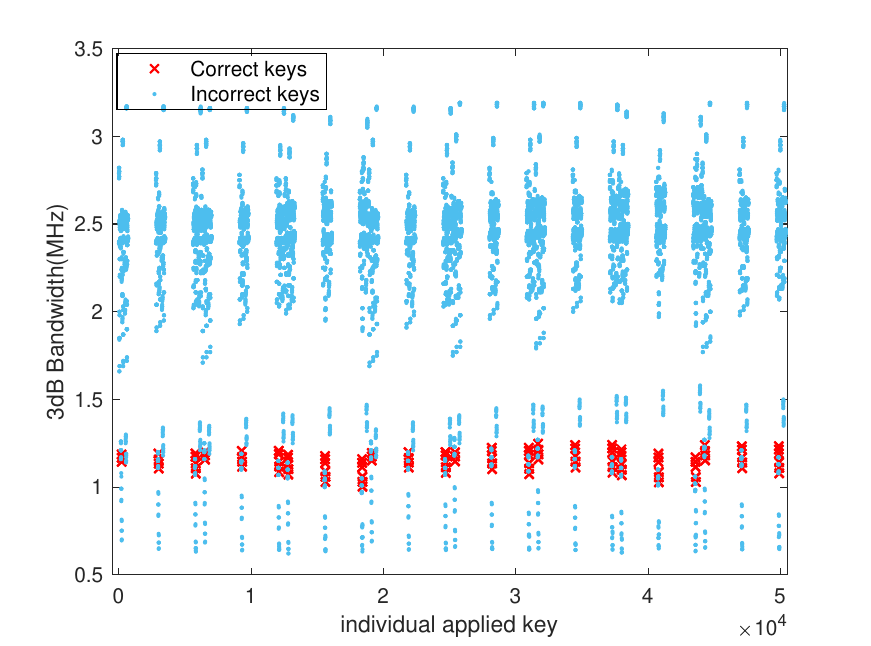}}
\caption{Variation in the 3dB bandwidth of the OTA for the applied keys.}
\label{bandwidth}
\end{figure}

\begin{figure}[t]
\centerline{\includegraphics[width=.75\linewidth]{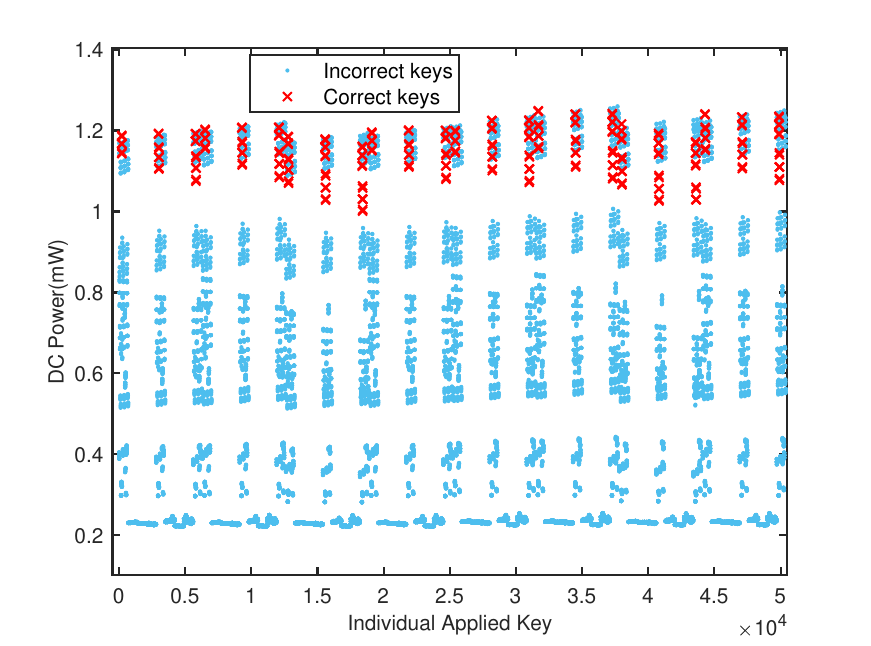}}
\caption{DC power consumption for the applied keys.}
\label{power}
\end{figure}

To demonstrate the robustness of the obfuscation achieved, we conducted simulations to evaluate the impact of $50400$ keys on the gain, phase, $3dB$ bandwidth (BW), and DC power of the OTA. Figure \ref{gain_phase} illustrates the effect of the keys on the gain, showing a wide range of degradation (up to $130dB$) due to different obfuscation arrangements. It can be observed that some keys result in a gain of $\geq70dB$, which meets the design specifications.
In this experiment, the rate of correct keys, which can be adjusted, accounts for $0.66\%$ of the total keys. Figure \ref{gain_phase} also displays a gap of $8dB$ between the plots, which is achieved by eliminating the nearly correct keys. This is accomplished by updating certain pairs of arrangements in the circuit. Furthermore, it is possible to remove the nearly correct keys that yield gain values between $67dB$ and $70dB$. However, these keys constitute less than $0.14\%$ of the total keys, indicating their negligible presence.
Overall, these simulation results highlight the effectiveness of the obfuscation technique in introducing significant variations in circuit performance across different keys, ensuring the robustness of the achieved obfuscation.
 Figure \ref{gain_phase} illustrates the impact of the applied keys on the phase, showing a degradation of up to $50$ degrees in the phase margin. Additionally, Figure \ref{bandwidth} and Figure \ref{power} present the impact of the applied keys on the 3dB bandwidth and DC power consumption, respectively.

The power consumption in the circuit for the correct keys falls within the range of $1.143mW$ to $1.188mW$. Interestingly, out of $1702$ keys that result in power consumption within this range, only $266$ of them are the correct keys. In other words, observing power consumption as a proxy for correctness might mislead an adversary. It is important to note that the simulation time for evaluating gain and power was approximately $55$ hours, highlighting the extensive computational effort involved in this analysis.
Furthermore, the introduction of $57$ added arrangements in the obfuscation process leads to a 1$58\%$ increase in circuit area. Moreover, power variations of up to $77\%$ were observed compared to the power consumed by the initial circuit, indicating the significant impact of obfuscation on power consumption.
These observations highlight the trade-offs and considerations involved in the obfuscation technique, including the impact on circuit performance, power consumption, and area overhead.
 
\begin{figure}[H]
\centerline{\includegraphics[scale=.55]{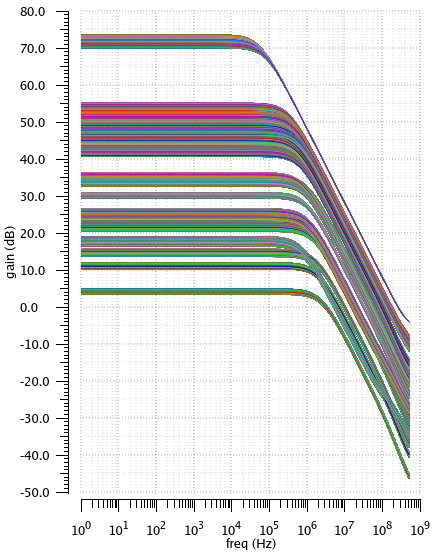}}
\caption{Layout-based effects on the OTA simulated for 300K keys.}
\label{gain2}
\end{figure}

In summary, we apply the following three techniques (and fourth one is discussed later on) to protect the correct keys and enhance the obfuscation process: 
\begin{enumerate}
\item Balance the effect of arrangements   
\item Remove pairs of arrangements producing a nearly correct performance
\item Remove pairs of arrangements with a relatively large impact on performance 
\end{enumerate}
 The third technique involves the elimination of pairs of arrangements that have a significant impact on the performance of the circuit. To better understand this technique, let us examine the circuit depicted in Figure \ref{OTA} and focus on transistors N7 and N8. When we choose non-symmetrical pairs of arrangements, such as BL-SP or SOD-BL, for N7 and N8, the circuit exhibits a negative gain, regardless of the other arrangements used. However, when we select symmetrical pairs of arrangements, such as SP-SP, BL-BL, or SOD-SOD, for N7 and N8, the gain becomes positive. We remove the non-symmetrical pairs of arrangements to eliminate the alarming effects on the circuit's performance and improve the quality of the obfuscation.
A point worth considering is that the application of these techniques may result in an uneven distribution of pairs of arrangements among different transistors, which can potentially raise concerns about the regularity of the circuit layout and unveil structural information. To address this issue, a possible solution is to equalize the number of pairs of arrangements for each transistor. Although this approach would reduce the keyspace, it would promote a more uniform layout and limit the disclosure of structural details.
 Indeed, the obfuscation technique employed in this work involves a trade-off between the key length and the level of output/behavior ``corruption" in the circuit, a concept also explored in digital logic locking \cite{b24}.
In order to expand the keyspace and address the issue of uneven pairs of arrangements, we consider five additional transistors: P16, P17, N13, N14, and N15. These transistors are specifically chosen to create additional pairs, enhancing the obfuscation of the circuit. By incorporating additional pairs of arrangements, the locked design now consists of a total of 31 pairs of arrangements. This is achieved by subtracting the six removed pairs from the original 28 pairs and adding nine new pairs. These 31 pairs of arrangements serve to hide eight pairs of transistors from the original base design. Furthermore, the single arrangement that was added in the previous experiment is still present in the locked design.
As a result, the keylength is now $41$ bits, achieved by adding $63$ arrangements to the original design. We simulated the circuit for $340,200$ keys. Figure \ref{gain2} demonstrates the impact of these keys on gain. The desired target keys account for less than $2\%$ of the overall keys. Importantly, all gains are now positive, with a minimum target value of $70dB$.
The simulation time for evaluating gain, $3dB$ bandwidth, and power consumption was approximately $22$ days. The circuit's area increased by $175\%$ due to $63$ added arrangements, and power variations of up to $73\%$ compared to the base circuit were observed. These simulations were performed on a server equipped with an Intel Xeon Gold $5122$ CPU with $32$ cores running @ $3.60GHz$.

The proposed locking scheme can be applied to larger analog circuits beyond the representative OTA block. It may not be necessary to apply the locking scheme to all analog blocks in a circuit. Once one block is locked, altering its performance is likely to affect the overall circuit performance, especially in multi-stage circuits. It should be noted that the overhead of the locking scheme for a single obfuscated transistor, in isolation, is $300\%$. However, for an entire circuit, the overhead will not be as significant since obfuscation can be applied selectively: It is important to highlight that not all transistors are obfuscated, as some may not be suitable candidates, and certain pairs of transistors are jointly obfuscated by fewer combinations of arrangements. While state-of-the-art approaches \cite {b12, b13} have achieved smaller overheads, they are also susceptible to SMT-based attacks \cite{b21}. In our approach, we strike a balance between overhead and security, prioritizing higher security. The security aspect of the locking scheme is further elaborated in Section~\ref{sec4}.

\section{Discussions}\label{sec4}

In our threat model, we consider both the foundry and the end-user as potentially untrusted entities. We assume that the foundry has complete knowledge/visibility over the IP except for the correct key(s). The malicious end-user is assumed to possess the necessary expertise and tools for reverse engineering the IP, including high-precision optical imaging equipment, circuit simulators, and functional copies of the IP as an oracle. However, the end-user does not have the level of visibility into the layout-dependent effects as it is typically not a current practice in RE efforts. It is also assumed that the end-user does not have access to a detailed transistor model that accounts for LDEs. Moreover, the end-user is aware of selecting only one arrangement for each transistor and not more than one.
In addition, we justify the inefficiency of several attack scenarios on the proposed approach, namely brute-force attack, SMT-based attack, and removal attack. We consider the following scenarios for attacking the proposed approach.

\subsection{Untrusted Foundry}\label{subsec41}
Everything about the design including LDE-level details is known to the foundry except for the correct key(s). To enhance the protection of the keys, we employ an additional technique, referred to as the fourth technique, in addition to the three techniques previously described in Section ~\ref{sec3}:

\begin{enumerate}
	\setcounter{enumi}{3}
	\item Making the order of the arrangements in the layout design arbitrary
\end{enumerate}
To further enhance the security of the keys, we applied an additional technique, which involves making the order of the arrangements in the layout design arbitrary. This means that the specific arrangement of the transistors in the layout is randomized, adding an extra layer of obfuscation to the design. By introducing this randomness, it becomes more difficult for an attacker, such as the foundry, to determine the correct arrangement and infer the corresponding keys. This method aims to thwart simple guesses made by the attacker, such as assuming that all arrangements follow a specific pattern (e.g., all arrangements are of a certain type, like BL). The arbitrary order of the arrangements introduces further complexity and unpredictability, making it harder for an attacker to reverse engineer the correct keys and compromise the security of the locked design. 
Given these considerations, we now address the following questions:

\hfill \break
\emph{Can a brute force attack compromise the design}? 
The key sizes used in the examples discussed are, technically, susceptible to brute force attacks, especially when the attack is mounted on a real device by observing its performance. However, it is important to note that the simulation time for evaluating the keys in the mentioned example was already significant, taking $22$ consecutive days to evaluate only $300K$ keys, which represents a very small subset of the potential keyspace. For larger circuits and longer keylengths, the computational requirements for a brute force attack become impractical and infeasible. Therefore, while the considered key sizes may be vulnerable to brute force attacks in certain scenarios, the time and resources required for such attacks increase significantly as the keylength and complexity of the circuit increase.

\hfill \break
\emph{Do partial simulations help to obtain the keys}? \emph{Or, in other words, can an adversary decompose the problem into smaller ones and apply a divide and conquer strategy}? Consider the input differential pairs in the OTA as an example. If an adversary attempts different combinations of arrangements for the transistors P1, P2, N1, and N2 to find a correct $g_{m}$, they may indeed find multiple combinations that yield the desired $g_{m}$ value. However, it is important to note that achieving the correct $g_{m}$ alone is not sufficient to unlock the circuit.
The circuit specifications involve multiple performance parameters beyond just $g_{m}$. While the adversary may find combinations that satisfy $g_{m}$, it is highly likely that most of these combinations will not meet the other required specifications of the circuit. To successfully unlock the circuit, the adversary would need to find keys that simultaneously satisfy all the desired specifications.
Expanding the search space to find keys that satisfy multiple specifications simultaneously significantly increases the complexity of the problem. It could easily lead to exploring a substantial portion, if not the entire keyspace. Moreover, the value of $g_{m}$ is dependent on the bias circuit, which is also obfuscated. Therefore, there might be incorrect bias values that still result in the desired $g_{m}$ value, further complicating the search for the correct key.
In summary, finding a key that deterministically satisfies multiple specifications at the same time is highly challenging. The search space is vast and the interdependencies between different specifications, as well as the obfuscation techniques employed, make it extremely difficult for an adversary to find the correct key solely by exploring different combinations of arrangements.

\begin{figure}[t]
\centerline{\includegraphics[width=.75\linewidth]{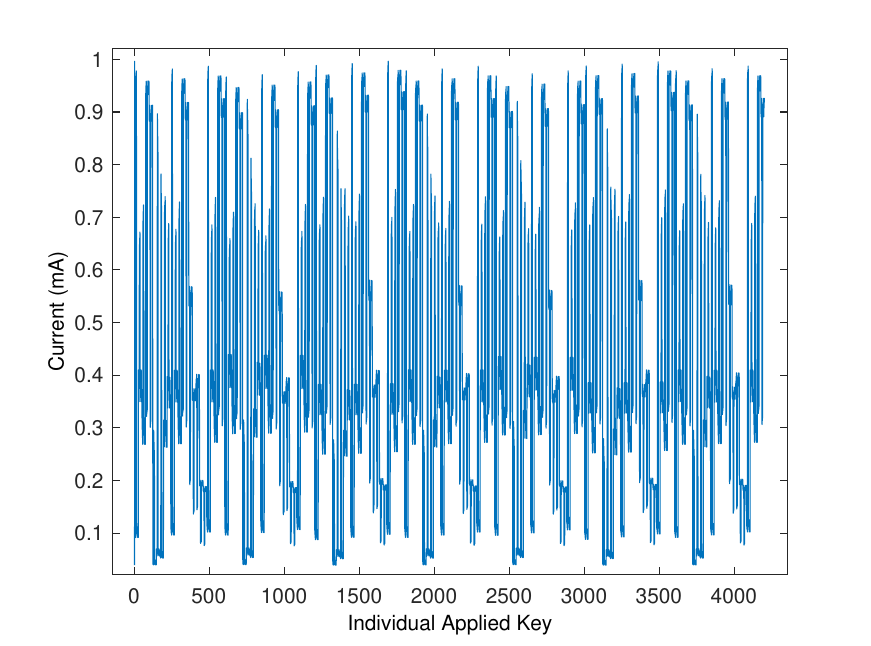}}
\caption{Current variations in an OTA branch for 4K different keys.}
\label{current_variations}
\end{figure}

\hfill \break
\emph{Is the SMT-based attack applicable to the proposed approach}? No. The SMT-based attack has been employed on analog ICs with locked bias circuits, where obfuscated current mirrors or voltage dividers are involved \cite{b12,b13}. In these cases, the correct key corresponds to a selection of mirrored branches with different transistor sizes, resulting in the desired sum of current. To find this selection, a simple equation is formulated, connecting the current of the reference branch to the currents of the mirrored branches, and the task is delegated to an SMT solver. The parameters necessary for this equation can be obtained from circuit specifications or the Process Design Kit (PDK) documentation. The SMT solver can solve this equation without relying on a circuit simulator.
This type of attack has also been applied to camouflaged analog IP \cite{b15} based on the same principle (Table ~\ref{comparisons}). However, in our approach, the layout-based effects are applied to multiple parts of the circuit, not just the bias circuit. Consequently, utilizing SMT-based attacks that target the bias circuit alone is insufficient for overcoming our approach. The equations that establish the link between the undesirable layout-based effects and circuit performance must be solved using a circuit simulator. This requirement presents \textbf{scalability challenges}, as the computational burden increases with the complexity and size of the circuit.
In summary, while SMT-based attacks have been successfully applied to certain analog ICs with locked bias circuits, our approach extends beyond the bias circuit and introduces layout-based effects to multiple parts of the circuit, including input differential pairs and summing circuit, as illustrated in Figure \ref{OTA}. Solving the equations that capture the impact of these effects on circuit performance necessitates the use of a circuit simulator, making the approach less scalable compared to the SMT-based attack.
 In Figure \ref{current_variations}, it is evident that there is a wide range of current variations observed in one branch of the OTA circuit. To effectively solve the equations using an SMT solver, the solver needs to be aware of the desirable range of currents in each branch. This information can only be obtained through extensive simulations. This poses a challenge as the existing SMT-based attack does not require such extensive simulations because the currents in those circuit equations are functions of fixed reference currents.

Furthermore, it is worth noting that a recent attack has been developed specifically targeting analog biasing locking techniques \cite{b22}. However, this attack focuses on searching for a correct bias instead of determining the key itself. Consequently, this attack is not applicable to our proposed technique, which obfuscates not only the bias circuit but also other parts of the circuit. By extending the obfuscation to multiple circuit components, our approach adds an extra layer of security and complexity, making it more challenging for attackers to extract the correct keys.

\hfill \break
\emph{Is the removal attack applicable to the proposed approach}? No. The removal attack aims to retrieve the base design by identifying and removing the protection circuitry \cite{b25}. However, in our locking scheme, the protected parts cannot be immediately distinguished from the original design, making it challenging to mount a successful removal attack. Since our method obfuscates multiple blocks, not just the biasing block, removing the key-bit transistors would require redesigning the entire circuit from scratch. In the case of the OTA, removing the key-bit transistors would eliminate approximately $50\%$ of the original design, rendering the attack ineffective.
In contrast, state-of-the-art techniques that focus solely on biasing blocks are vulnerable to removal attacks \cite{b11,b12,b13,b16}. In such cases, the attacker only needs to recover the biasing blocks, which typically consist of a small number of transistors. Similarly, locked AMS designs in other approaches can also be vulnerable to removal attacks by removing the digital lock and redesigning the small biasing blocks \cite{b18,b19}.
Table \ref{comparisons} summarizes the security-overhead trade-off achieved by our approach compared to other techniques. Our approach establishes a balance between security and overhead, with the ability to reduce the area overhead to approximately $30\%$ by selecting two arrangements per obfuscated transistor instead of three. However, it is important to note that lowering the number of arrangements per transistor would also lower the security level of the locked circuit.

\begin{sidewaystable}[]
	\caption{Vulnerability of state-of-the-art DfTr methods to SMT-based attack}\label{comparisons}
	\begin{center}
		\begin{tabular}{c c c c c }
			\hline
			\hline
			 \multirow{2}{*}{\sc DfTr technique} & {\sc Susceptible to} & {\sc Susceptible to} & \multirow{2}{*}{\sc Purely analog} & \multirow{2}{*} {\sc Area overhead (\%)}  \\
			 & {\sc SMT-based attack} & {\small\sc removal attack}\\
		    \hline
		    \hline
            \cite{b11} & Yes & Yes &No & $-$ \\
			\hline
            \cite{b12} & Yes & Yes & Yes & 6.3  \\
			\hline
            \cite{b13} & Yes & Yes & Yes & 6.64    \\
			\hline
            \cite{b15} & Yes & No &Yes & up to 48*   \\
			\hline
            \cite{b16} & Yes & Yes & Yes & $-$   \\
			\hline
            \cite{b18} & No  & Yes &No &  0 $\sim$ 171.3**\\
			\hline
            \cite{b19} & No    & Yes &No & 6.7 $\sim$ 24.4** \\
			\hline
			\textbf{This work} & No & No & Yes & 30.6 $\sim$ 175***\\
			\hline
			\hline
		\end{tabular}
		\vspace{2pt}
		\item *This is not a key-based technique, thus the relatively low overhead.
		\item **These values vary depending on the obfuscated circuit and parameters of the locking scheme.
		\item ***Depending on the number of arrangements per transistors selected for obfuscation, which is either 2 or 3, the area overhead varies as shown above. 
	\end{center}
	
\end{sidewaystable}

\subsection{Untrusted End-user}\label{subsec42}

In the scenario where the netlist of the locked circuit is obtained through reverse engineering efforts, the adversary will have access to the metal lines, vias, contacts, and poly lines of the circuit. However, it is important to note that the adversary does not have access to LDE-level visibility, which means they cannot observe the detailed characteristics and behavior of the transistors.
Upon obtaining the locked netlist, the adversary will observe groups of transistors with identical sizes, representing the arrangements used in our locking scheme. However, since we do not manipulate the transistor's width (W) or length (L), the adversary's model will not capture the layout-dependent effects that were originally designed to exploit. Therefore, simulating the obtained netlist with different keys will result in the same behavior, which is incorrect if the circuit was specifically designed to utilize LDEs.
Even if the adversary has access to an oracle that can confirm that different keys lead to different performance, they have no means to map these performance variations back to the circuit's design. This lack of detailed knowledge about the LDEs prevents the adversary from establishing useful distinguishing input patterns, similar to the SAT attack \cite{b26}. 
Consequently, the adversary's chances of unlocking the circuit are not higher than those of a malicious foundry, even when they possess an oracle.

Along similar lines, the genetic algorithm-based attack \cite{b27}, which relies on oracle and locked netlist, is unlikely to be effective against the proposed approach. The genetic algorithm-based attack utilizes evolutionary search techniques to explore the design space and find potential keys that unlock the circuit. However, since our threat model does not provide access to the detailed locked netlist, the adversary lacks the necessary information to conduct such an attack.
Without access to the detailed locked netlist, the adversary is unable to accurately model the circuit's behavior and the impact of different keys on its performance. This lack of detailed information about the circuit's design and layout-dependent effects makes it extremely challenging for the adversary to successfully apply the genetic algorithm-based attack to reverse engineer or unlock the locked analog ICs.

In summary, the absence of LDE-awareness in the netlist obtained through reverse engineering makes it extremely challenging for the adversary to accurately understand and exploit the design's key-dependent performance variations, thus impeding their ability to unlock the circuit.

\section{Conclusion}\label{sec5}

This paper presents a novel approach for locking analog integrated circuits by leveraging layout-based effects such as Well Proximity Effect and Length of Oxide Diffusion. The proposed approach is demonstrated on an operational transconductance amplifier circuit using a large number of keys to showcase the effectiveness of the obfuscation achieved.
By applying the layout-based effects to the circuit, we show that the gain, phase margin, 3dB bandwidth, and power characteristics are significantly altered, thereby enhancing the security of the locked circuit. These layout-based effects serve as a form of obfuscation, making it difficult for adversaries to reverse engineer or counterfeit the circuit.
The results of this work demonstrate the potential of the proposed approach in protecting analog circuits against counterfeiting and reverse engineering attacks, which are common threats in the semiconductor industry.

As a future direction, we plan to validate the methodology in silicon by utilizing a commercial foundry service. This step will provide a realistic scenario of outsourcing, where the circuit is fabricated by a third-party foundry. By implementing the proposed approach in silicon, the authors can evaluate its practicality, performance, and effectiveness in a real-world setting.

\bmhead{Acknowledgments}
This work has received partial funding from the European Union through the European Social Fund as part of the ``ICT programme". Additionally, it has also been supported by the European Union's Horizon 2020 research and innovation programme under grant agreement No 952252 (SAFEST). 

\bibliography{sn-bibliography}%


\begin{thebibliography}{29}
\ifx \bisbn   \undefined \def \bisbn  #1{ISBN #1}\fi
\ifx \binits  \undefined \def \binits#1{#1}\fi
\ifx \bauthor  \undefined \def \bauthor#1{#1}\fi
\ifx \batitle  \undefined \def \batitle#1{#1}\fi
\ifx \bjtitle  \undefined \def \bjtitle#1{#1}\fi
\ifx \bvolume  \undefined \def \bvolume#1{\textbf{#1}}\fi
\ifx \byear  \undefined \def \byear#1{#1}\fi
\ifx \bissue  \undefined \def \bissue#1{#1}\fi
\ifx \bfpage  \undefined \def \bfpage#1{#1}\fi
\ifx \blpage  \undefined \def \blpage #1{#1}\fi
\ifx \burl  \undefined \def \burl#1{\textsf{#1}}\fi
\ifx \doiurl  \undefined \def \doiurl#1{\url{https://doi.org/#1}}\fi
\ifx \betal  \undefined \def \betal{\textit{et al.}}\fi
\ifx \binstitute  \undefined \def \binstitute#1{#1}\fi
\ifx \binstitutionaled  \undefined \def \binstitutionaled#1{#1}\fi
\ifx \bctitle  \undefined \def \bctitle#1{#1}\fi
\ifx \beditor  \undefined \def \beditor#1{#1}\fi
\ifx \bpublisher  \undefined \def \bpublisher#1{#1}\fi
\ifx \bbtitle  \undefined \def \bbtitle#1{#1}\fi
\ifx \bedition  \undefined \def \bedition#1{#1}\fi
\ifx \bseriesno  \undefined \def \bseriesno#1{#1}\fi
\ifx \blocation  \undefined \def \blocation#1{#1}\fi
\ifx \bsertitle  \undefined \def \bsertitle#1{#1}\fi
\ifx \bsnm \undefined \def \bsnm#1{#1}\fi
\ifx \bsuffix \undefined \def \bsuffix#1{#1}\fi
\ifx \bparticle \undefined \def \bparticle#1{#1}\fi
\ifx \barticle \undefined \def \barticle#1{#1}\fi
\bibcommenthead
\ifx \bconfdate \undefined \def \bconfdate #1{#1}\fi
\ifx \botherref \undefined \def \botherref #1{#1}\fi
\ifx \url \undefined \def \url#1{\textsf{#1}}\fi
\ifx \bchapter \undefined \def \bchapter#1{#1}\fi
\ifx \bbook \undefined \def \bbook#1{#1}\fi
\ifx \bcomment \undefined \def \bcomment#1{#1}\fi
\ifx \oauthor \undefined \def \oauthor#1{#1}\fi
\ifx \citeauthoryear \undefined \def \citeauthoryear#1{#1}\fi
\ifx \endbibitem  \undefined \def \endbibitem {}\fi
\ifx \bconflocation  \undefined \def \bconflocation#1{#1}\fi
\ifx \arxivurl  \undefined \def \arxivurl#1{\textsf{#1}}\fi
\csname PreBibitemsHook\endcsname

\bibitem[\protect\citeauthoryear{Bhasin and Regazzoni}{2015}]{b1}
\begin{bchapter}
\bauthor{\bsnm{Bhasin}, \binits{S.}},
\bauthor{\bsnm{Regazzoni}, \binits{F.}}:
\bctitle{A survey on hardware trojan detection techniques}.
In: \bbtitle{2015 IEEE International Symposium on Circuits and Systems
  (ISCAS)},
pp. \bfpage{2021}--\blpage{2024}
(\byear{2015}).
\doiurl{10.1109/ISCAS.2015.7169073}
\end{bchapter}
\endbibitem

\bibitem[\protect\citeauthoryear{Dupuis et~al.}{2014}]{b2}
\begin{bchapter}
\bauthor{\bsnm{Dupuis}, \binits{S.}},
\bauthor{\bsnm{Ba}, \binits{P.-S.}},
\bauthor{\bsnm{Di~Natale}, \binits{G.}},
\bauthor{\bsnm{Flottes}, \binits{M.-L.}},
\bauthor{\bsnm{Rouzeyre}, \binits{B.}}:
\bctitle{A novel hardware logic encryption technique for thwarting illegal
  overproduction and hardware trojans}.
In: \bbtitle{2014 IEEE 20th International On-Line Testing Symposium (IOLTS)},
pp. \bfpage{49}--\blpage{54}
(\byear{2014}).
\doiurl{10.1109/IOLTS.2014.6873671}
\end{bchapter}
\endbibitem

\bibitem[\protect\citeauthoryear{Jacob et~al.}{2014}]{b3}
\begin{barticle}
\bauthor{\bsnm{Jacob}, \binits{N.}},
\bauthor{\bsnm{Merli}, \binits{D.}},
\bauthor{\bsnm{Heyszl}, \binits{J.}},
\bauthor{\bsnm{Sigl}, \binits{G.}}:
\batitle{Hardware trojans: current challenges and approaches}.
\bjtitle{IET Computers \& Digital Techniques}
\bvolume{8}(\bissue{6}),
\bfpage{264}--\blpage{273}
(\byear{2014})
\end{barticle}
\endbibitem

\bibitem[\protect\citeauthoryear{Keshavarz et~al.}{2018}]{b4}
\begin{barticle}
\bauthor{\bsnm{Keshavarz}, \binits{S.}},
\bauthor{\bsnm{Yu}, \binits{C.}},
\bauthor{\bsnm{Ghandali}, \binits{S.}},
\bauthor{\bsnm{Xu}, \binits{X.}},
\bauthor{\bsnm{Holcomb}, \binits{D.}}:
\batitle{Survey on applications of formal methods in reverse engineering and
  intellectual property protection}.
\bjtitle{Journal of Hardware and Systems Security}
\bvolume{2}(\bissue{3}),
\bfpage{214}--\blpage{224}
(\byear{2018})
\end{barticle}
\endbibitem

\bibitem[\protect\citeauthoryear{SEMI}{2008}]{b5}
\begin{botherref}
\oauthor{\bsnm{SEMI}}:
{White paper: Innovation at Risk — Intellectual Property Challenges and
  Opportunities}.
Technical report
(April 2008)
\end{botherref}
\endbibitem

\bibitem[\protect\citeauthoryear{Colombier and Bossuet}{2014}]{b6}
\begin{barticle}
\bauthor{\bsnm{Colombier}, \binits{B.}},
\bauthor{\bsnm{Bossuet}, \binits{L.}}:
\batitle{Survey of hardware protection of design data for integrated circuits
  and intellectual properties}.
\bjtitle{IET Computers \& Digital Techniques}
\bvolume{8}(\bissue{6}),
\bfpage{274}--\blpage{287}
(\byear{2014})
\end{barticle}
\endbibitem

\bibitem[\protect\citeauthoryear{Rajendran et~al.}{2014}]{b7}
\begin{barticle}
\bauthor{\bsnm{Rajendran}, \binits{J.}},
\bauthor{\bsnm{Sinanoglu}, \binits{O.}},
\bauthor{\bsnm{Karri}, \binits{R.}}:
\batitle{Regaining trust in vlsi design: Design-for-trust techniques}.
\bjtitle{Proceedings of the IEEE}
\bvolume{102}(\bissue{8}),
\bfpage{1266}--\blpage{1282}
(\byear{2014})
\end{barticle}
\endbibitem

\bibitem[\protect\citeauthoryear{Roy et~al.}{2010}]{b8}
\begin{barticle}
\bauthor{\bsnm{Roy}, \binits{J.A.}},
\bauthor{\bsnm{Koushanfar}, \binits{F.}},
\bauthor{\bsnm{Markov}, \binits{I.L.}}:
\batitle{Ending piracy of integrated circuits}.
\bjtitle{Computer}
\bvolume{43}(\bissue{10}),
\bfpage{30}--\blpage{38}
(\byear{2010})
\end{barticle}
\endbibitem

\bibitem[\protect\citeauthoryear{Sanabria-Borbon et~al.}{2020}]{b9}
\begin{bchapter}
\bauthor{\bsnm{Sanabria-Borbon}, \binits{A.}},
\bauthor{\bsnm{Jayasankaran}, \binits{N.G.}},
\bauthor{\bsnm{Lee}, \binits{S.}},
\bauthor{\bsnm{S{\'a}nchez-Sinencio}, \binits{E.}},
\bauthor{\bsnm{Hu}, \binits{J.}},
\bauthor{\bsnm{Rajendran}, \binits{J.}}:
\bctitle{Schmitt trigger-based key provisioning for locking analog/rf
  integrated circuits}.
In: \bbtitle{2020 IEEE International Test Conference (ITC)},
pp. \bfpage{1}--\blpage{10}
(\byear{2020}).
\bcomment{IEEE}
\end{bchapter}
\endbibitem

\bibitem[\protect\citeauthoryear{Elshamy et~al.}{2020}]{b10}
\begin{bchapter}
\bauthor{\bsnm{Elshamy}, \binits{M.}},
\bauthor{\bsnm{Sayed}, \binits{A.}},
\bauthor{\bsnm{Lou{\"e}rat}, \binits{M.-M.}},
\bauthor{\bsnm{Rhouni}, \binits{A.}},
\bauthor{\bsnm{Aboushady}, \binits{H.}},
\bauthor{\bsnm{Stratigopoulos}, \binits{H.-G.}}:
\bctitle{Securing programmable analog ics against piracy}.
In: \bbtitle{2020 Design, Automation \& Test in Europe Conference \& Exhibition
  (DATE)},
pp. \bfpage{61}--\blpage{66}
(\byear{2020}).
\bcomment{IEEE}
\end{bchapter}
\endbibitem

\bibitem[\protect\citeauthoryear{Hoe et~al.}{2014}]{b11}
\begin{bchapter}
\bauthor{\bsnm{Hoe}, \binits{D.H.}},
\bauthor{\bsnm{Rajendran}, \binits{J.}},
\bauthor{\bsnm{Karri}, \binits{R.}}:
\bctitle{Towards secure analog designs: A secure sense amplifier using
  memristors}.
In: \bbtitle{2014 IEEE Computer Society Annual Symposium on VLSI},
pp. \bfpage{516}--\blpage{521}
(\byear{2014}).
\bcomment{IEEE}
\end{bchapter}
\endbibitem

\bibitem[\protect\citeauthoryear{Rao and Savidis}{2017}]{b12}
\begin{bchapter}
\bauthor{\bsnm{Rao}, \binits{V.V.}},
\bauthor{\bsnm{Savidis}, \binits{I.}}:
\bctitle{Protecting analog circuits with parameter biasing obfuscation}.
In: \bbtitle{2017 18th IEEE Latin American Test Symposium (LATS)},
pp. \bfpage{1}--\blpage{6}
(\byear{2017}).
\bcomment{IEEE}
\end{bchapter}
\endbibitem

\bibitem[\protect\citeauthoryear{Wang et~al.}{2017}]{b13}
\begin{bchapter}
\bauthor{\bsnm{Wang}, \binits{J.}},
\bauthor{\bsnm{Shi}, \binits{C.}},
\bauthor{\bsnm{Sanabria-Borbon}, \binits{A.}},
\bauthor{\bsnm{S{\'a}nchez-Sinencio}, \binits{E.}},
\bauthor{\bsnm{Hu}, \binits{J.}}:
\bctitle{Thwarting analog ic piracy via combinational locking}.
In: \bbtitle{2017 IEEE International Test Conference (ITC)},
pp. \bfpage{1}--\blpage{10}
(\byear{2017}).
\bcomment{IEEE}
\end{bchapter}
\endbibitem

\bibitem[\protect\citeauthoryear{Nimmalapudi et~al.}{2020}]{b14}
\begin{bchapter}
\bauthor{\bsnm{Nimmalapudi}, \binits{S.G.R.}},
\bauthor{\bsnm{Volanis}, \binits{G.}},
\bauthor{\bsnm{Lu}, \binits{Y.}},
\bauthor{\bsnm{Antonopoulos}, \binits{A.}},
\bauthor{\bsnm{Marshall}, \binits{A.}},
\bauthor{\bsnm{Makris}, \binits{Y.}}:
\bctitle{Range-controlled floating-gate transistors: A unified solution for
  unlocking and calibrating analog ics}.
In: \bbtitle{2020 Design, Automation \& Test in Europe Conference \& Exhibition
  (DATE)},
pp. \bfpage{286}--\blpage{289}
(\byear{2020}).
\bcomment{IEEE}
\end{bchapter}
\endbibitem

\bibitem[\protect\citeauthoryear{Ash-Saki and Ghosh}{2018}]{b15}
\begin{bchapter}
\bauthor{\bsnm{Ash-Saki}, \binits{A.}},
\bauthor{\bsnm{Ghosh}, \binits{S.}}:
\bctitle{How multi-threshold designs can protect analog ips}.
In: \bbtitle{2018 IEEE 36th International Conference on Computer Design
  (ICCD)},
pp. \bfpage{464}--\blpage{471}
(\byear{2018}).
\bcomment{IEEE}
\end{bchapter}
\endbibitem

\bibitem[\protect\citeauthoryear{Volanis et~al.}{2019}]{b16}
\begin{bchapter}
\bauthor{\bsnm{Volanis}, \binits{G.}},
\bauthor{\bsnm{Lu}, \binits{Y.}},
\bauthor{\bsnm{Nimmalapudi}, \binits{S.G.R.}},
\bauthor{\bsnm{Antonopoulos}, \binits{A.}},
\bauthor{\bsnm{Marshall}, \binits{A.}},
\bauthor{\bsnm{Makris}, \binits{Y.}}:
\bctitle{Analog performance locking through neural network-based biasing}.
In: \bbtitle{2019 IEEE 37th VLSI Test Symposium (VTS)},
pp. \bfpage{1}--\blpage{6}
(\byear{2019}).
\bcomment{IEEE}
\end{bchapter}
\endbibitem

\bibitem[\protect\citeauthoryear{Elshamy et~al.}{2020}]{b17}
\begin{bchapter}
\bauthor{\bsnm{Elshamy}, \binits{M.}},
\bauthor{\bsnm{Sayed}, \binits{A.}},
\bauthor{\bsnm{Lou{\"e}rat}, \binits{M.-M.}},
\bauthor{\bsnm{Rhouni}, \binits{A.}},
\bauthor{\bsnm{Aboushady}, \binits{H.}},
\bauthor{\bsnm{Stratigopoulos}, \binits{H.-G.}}:
\bctitle{Securing programmable analog ics against piracy}.
In: \bbtitle{2020 Design, Automation \& Test in Europe Conference \& Exhibition
  (DATE)},
pp. \bfpage{61}--\blpage{66}
(\byear{2020}).
\bcomment{IEEE}
\end{bchapter}
\endbibitem

\bibitem[\protect\citeauthoryear{Tlili et~al.}{2022}]{b28}
\begin{bchapter}
\bauthor{\bsnm{Tlili}, \binits{M.}},
\bauthor{\bsnm{Sayed}, \binits{A.}},
\bauthor{\bsnm{Mahmoud}, \binits{D.}},
\bauthor{\bsnm{Lou{\"e}rat}, \binits{M.-M.}},
\bauthor{\bsnm{Aboushady}, \binits{H.}},
\bauthor{\bsnm{Stratigopoulos}, \binits{H.-G.}}:
\bctitle{Anti-piracy of analog and mixed-signal circuits in fd-soi}.
In: \bbtitle{2022 27th Asia and South Pacific Design Automation Conference
  (ASP-DAC)},
pp. \bfpage{423}--\blpage{428}
(\byear{2022}).
\bcomment{IEEE}
\end{bchapter}
\endbibitem

\bibitem[\protect\citeauthoryear{Jayasankaran et~al.}{2018}]{b18}
\begin{bchapter}
\bauthor{\bsnm{Jayasankaran}, \binits{N.G.}},
\bauthor{\bsnm{Borbon}, \binits{A.S.}},
\bauthor{\bsnm{Sanchez-Sinencio}, \binits{E.}},
\bauthor{\bsnm{Hu}, \binits{J.}},
\bauthor{\bsnm{Rajendran}, \binits{J.}}:
\bctitle{Towards provably-secure analog and mixed-signal locking against
  overproduction}.
In: \bbtitle{Proceedings of the International Conference on Computer-Aided
  Design},
pp. \bfpage{1}--\blpage{8}
(\byear{2018})
\end{bchapter}
\endbibitem

\bibitem[\protect\citeauthoryear{Leonhard et~al.}{2019}]{b19}
\begin{bchapter}
\bauthor{\bsnm{Leonhard}, \binits{J.}},
\bauthor{\bsnm{Yasin}, \binits{M.}},
\bauthor{\bsnm{Turk}, \binits{S.}},
\bauthor{\bsnm{Nabeel}, \binits{M.T.}},
\bauthor{\bsnm{Lou{\"e}rat}, \binits{M.-M.}},
\bauthor{\bsnm{Chotin-Avot}, \binits{R.}},
\bauthor{\bsnm{Aboushady}, \binits{H.}},
\bauthor{\bsnm{Sinanoglu}, \binits{O.}},
\bauthor{\bsnm{Stratigopoulos}, \binits{H.-G.}}:
\bctitle{Mixlock: Securing mixed-signal circuits via logic locking}.
In: \bbtitle{2019 Design, Automation \& Test in Europe Conference \& Exhibition
  (DATE)},
pp. \bfpage{84}--\blpage{89}
(\byear{2019}).
\bcomment{IEEE}
\end{bchapter}
\endbibitem

\bibitem[\protect\citeauthoryear{Rao et~al.}{2020}]{b20}
\begin{bchapter}
\bauthor{\bsnm{Rao}, \binits{V.V.}},
\bauthor{\bsnm{Juretus}, \binits{K.}},
\bauthor{\bsnm{Savidis}, \binits{I.}}:
\bctitle{Security vulnerabilities of obfuscated analog circuits}.
In: \bbtitle{2020 IEEE International Symposium on Circuits and Systems
  (ISCAS)},
pp. \bfpage{1}--\blpage{5}
(\byear{2020}).
\bcomment{IEEE}
\end{bchapter}
\endbibitem

\bibitem[\protect\citeauthoryear{Jayasankaran et~al.}{2020}]{b21}
\begin{barticle}
\bauthor{\bsnm{Jayasankaran}, \binits{N.G.}},
\bauthor{\bsnm{Sanabria-Borb{\'o}n}, \binits{A.}},
\bauthor{\bsnm{Abuellil}, \binits{A.}},
\bauthor{\bsnm{S{\'a}nchez-Sinencio}, \binits{E.}},
\bauthor{\bsnm{Hu}, \binits{J.}},
\bauthor{\bsnm{Rajendran}, \binits{J.}}:
\batitle{Breaking analog locking techniques}.
\bjtitle{IEEE Transactions on Very Large Scale Integration (VLSI) Systems}
\bvolume{28}(\bissue{10}),
\bfpage{2157}--\blpage{2170}
(\byear{2020})
\end{barticle}
\endbibitem

\bibitem[\protect\citeauthoryear{Leonhard et~al.}{2021}]{b22}
\begin{bchapter}
\bauthor{\bsnm{Leonhard}, \binits{J.}},
\bauthor{\bsnm{Elshamy}, \binits{M.}},
\bauthor{\bsnm{Lou{\"e}rat}, \binits{M.-M.}},
\bauthor{\bsnm{Stratigopoulos}, \binits{H.-G.}}:
\bctitle{Breaking analog biasing locking techniques via re-synthesis}.
In: \bbtitle{Proceedings of the 26th Asia and South Pacific Design Automation
  Conference},
pp. \bfpage{555}--\blpage{560}
(\byear{2021})
\end{bchapter}
\endbibitem

\bibitem[\protect\citeauthoryear{Acharya et~al.}{2020}]{b27}
\begin{bchapter}
\bauthor{\bsnm{Acharya}, \binits{R.Y.}},
\bauthor{\bsnm{Chowdhury}, \binits{S.}},
\bauthor{\bsnm{Ganji}, \binits{F.}},
\bauthor{\bsnm{Forte}, \binits{D.}}:
\bctitle{Attack of the genes: Finding keys and parameters of locked analog ics
  using genetic algorithm}.
In: \bbtitle{2020 IEEE International Symposium on Hardware Oriented Security
  and Trust (HOST)},
pp. \bfpage{284}--\blpage{294}
(\byear{2020}).
\bcomment{IEEE}
\end{bchapter}
\endbibitem

\bibitem[\protect\citeauthoryear{Aljafar et~al.}{2022}]{b29}
\begin{bchapter}
\bauthor{\bsnm{Aljafar}, \binits{M.J.}},
\bauthor{\bsnm{Aza\"{\i}s}, \binits{F.}},
\bauthor{\bsnm{Flottes}, \binits{M.-L.}},
\bauthor{\bsnm{Pagliarini}, \binits{S.}}:
\bctitle{Leveraging layout-based effects for locking analog ics}.
In: \bbtitle{Proceedings of the 2022 Workshop on Attacks and Solutions in
  Hardware Security}.
\bsertitle{ASHES'22},
pp. \bfpage{5}--\blpage{13}.
\bpublisher{Association for Computing Machinery},
\blocation{New York, NY, USA}
(\byear{2022}).
\doiurl{10.1145/3560834.3563826}
\end{bchapter}
\endbibitem

\bibitem[\protect\citeauthoryear{{Intelligence Advanced Research Projects
  Activity (IARPA)}}{2016}]{b23}
\begin{botherref}
\oauthor{\bsnm{{Intelligence Advanced Research Projects Activity (IARPA)}}}:
Rapid Analysis of Various Emerging Nanoelectronics (RAVEN).
\url{https://www.iarpa.gov/index.php/research-programs/raven}
\end{botherref}
\endbibitem

\bibitem[\protect\citeauthoryear{Rajendran et~al.}{2012}]{b24}
\begin{bchapter}
\bauthor{\bsnm{Rajendran}, \binits{J.}},
\bauthor{\bsnm{Pino}, \binits{Y.}},
\bauthor{\bsnm{Sinanoglu}, \binits{O.}},
\bauthor{\bsnm{Karri}, \binits{R.}}:
\bctitle{Logic encryption: A fault analysis perspective}.
In: \bbtitle{2012 Design, Automation \& Test in Europe Conference \& Exhibition
  (DATE)},
pp. \bfpage{953}--\blpage{958}
(\byear{2012}).
\bcomment{IEEE}
\end{bchapter}
\endbibitem

\bibitem[\protect\citeauthoryear{Yasin et~al.}{2017}]{b25}
\begin{barticle}
\bauthor{\bsnm{Yasin}, \binits{M.}},
\bauthor{\bsnm{Mazumdar}, \binits{B.}},
\bauthor{\bsnm{Sinanoglu}, \binits{O.}},
\bauthor{\bsnm{Rajendran}, \binits{J.}}:
\batitle{Removal attacks on logic locking and camouflaging techniques}.
\bjtitle{IEEE Transactions on Emerging Topics in Computing}
\bvolume{8}(\bissue{2}),
\bfpage{517}--\blpage{532}
(\byear{2017})
\end{barticle}
\endbibitem

\bibitem[\protect\citeauthoryear{Subramanyan et~al.}{2015}]{b26}
\begin{bchapter}
\bauthor{\bsnm{Subramanyan}, \binits{P.}},
\bauthor{\bsnm{Ray}, \binits{S.}},
\bauthor{\bsnm{Malik}, \binits{S.}}:
\bctitle{Evaluating the security of logic encryption algorithms}.
In: \bbtitle{2015 IEEE International Symposium on Hardware Oriented Security
  and Trust (HOST)},
pp. \bfpage{137}--\blpage{143}
(\byear{2015}).
\bcomment{IEEE}
\end{bchapter}
\endbibitem

\end{thebibliography}

\end{document}